\documentclass[preprint,prc,showpacs,preprintnumbers,
               superscriptaddress,amsmath,amssymb,floatfix]{revtex4}

\usepackage{graphicx}
\usepackage{dcolumn}
\usepackage{bm}
\usepackage{longtable}

\newcommand{\beq}{\begin{equation}}
\newcommand{\eeq}{\end{equation}}
\newcommand{\beqn}{\begin{eqnarray}}
\newcommand{\eeqn}{\end{eqnarray}}
\newcommand{\btab}{\begin{tabular}}
\newcommand{\etab}{\end{tabular}}

\newcommand{\re}{\nonumber\\}

\newcommand{\lc}{\left<}
\newcommand{\rc}{\right>}
\newcommand{\lr}{\left|}
\newcommand{\rl}{\right|}
\newcommand{\lb}{\left(}
\newcommand{\rb}{\right)}
\newcommand{\ls}{\left[}
\newcommand{\rs}{\right]}
\newcommand{\Lb}{\left\{}
\newcommand{\Rb}{\right\}}

\begin{document}

\title{Description of Chiral Doublets in $A\sim130$ Nuclei and the Possible Chiral Doublets in $A\sim100$ Nuclei}
\author{J. Peng}
\affiliation{School of Physics, Peking University, Beijing 100871}
\author{J. Meng}\thanks{e-mail: mengj@pku.edu.cn}
\affiliation{School of Physics, Peking University, Beijing 100871}
\affiliation{Institute of Theoretical Physics, Chinese Academy of
Science, Beijing 100080}
\affiliation{Center of Theoretical Nuclear Physics, National Laboratory of \\
       Heavy Ion Accelerator, Lanzhou 730000}

\author{S. Q. Zhang}
\affiliation{School of Physics, Peking University, Beijing 100871}

\date{\today}

\begin{abstract}
The chiral doublets for nuclei in $A\sim100$ and $A\sim130$
regions have been studied with the particle-rotor model. The
experimental spectra of chiral partners bands for four $N=75$
isotones in $A\sim130$ region have been well reproduced by the
calculation with the configuration $\pi h_{11/2}\otimes\nu
h_{11/2}^{-1}$. The possible chiral doublets in $A\sim100$ region
have been predicted by the PRM model with the configuration $\pi
g_{9/2}\otimes\nu g_{9/2}^{-1}$ based on the analysis of the
spectra, the $\omega$-I relation, the $B(M1)$ and $B(E2)$
transition probabilities, and the structure of the wave functions.
The possible chiral doublets for nuclei with asymmetric particle
and hole configuration are also discussed.

\end{abstract}


\maketitle

\section{Introduction}

Static chiral symmetries exist commonly in nature. We can find
many examples including the macroscopic spirals of snail shells
and the microscopic handedness of certain molecules. In particle
physics, it is a dynamic property distinguishing between the
parallel and antiparallel orientations of the intrinsic spin with
respect to the momentum of the particle. In nuclear physics,
triaxially deformed doubly odd nuclei can rotate in a left-handed
and right-handed geometrical configuration, which is the
manifestation of chiral symmetry breaking predicted recently
\cite{Meng97}. The angular momenta of the core and of the odd
particles can form either a left-handed or a right-handed
combination. These two possibilities are transformed into each
other by the chiral operator which combines time reversal and
rotation by 180$^\circ$, $\chi={\cal{TR}}(\pi)$, instead of any
other simple rotation. In the ideal case, the two equivalent
chiral arrangements are independent with each other, and we can
get two degenerate bands with the same quasiparticle
configuration.

In Ref. \cite{Meng97}, a pair of $\bigtriangleup I=1$ bands found
in $^{134}$Pr ($N=75$), with the $\pi h_{11/2}\otimes\nu
h_{11/2}^{-1}$ configuration \cite{Petrache96}, has been suggested
as a candidate for the chiral doubling. Recently, the possible
experimental evidences are reported for a series of chiral-twin
bands in odd-$Z$ $N=75$ isotones, $^{130}$Cs, $^{132}$La,
$^{134}$Pr, $^{136}$Pm \cite{Starosta01} and $N=73$ isotones,
$^{128}$Cs, $^{130}$La, $^{132}$Pr \cite{Koike01}. Hartley
$et~al$. \cite{Hartley01} reported the chiral doublet bands in
$^{136}$Pm, and compared the calculated transition strength ratios
$B(M1)/B(E2)$ of the chiral bands with the experimental values for
the first time. Using a phenomenological core-particle-hole
coupling model similar as in Ref.[1], the chirality in odd-odd
nuclei has been studied based on the configuration $\pi
h_{11/2}\otimes\nu h_{11/2}^{-1}$ and compared with the
experimental data, more details can be found in
\cite{Starosta02,StarostaNPA01,Koike03} and the references
therein. In the mass region $A\sim130$, there are many other
experimental evidences \cite{Hecht01, Bark01, Petrache02, Li XF02}
which support the existence of the chiral doublet bands in
triaxially deformed doubly odd nuclei. Furthermore, the first
chiral bands in the even-even nucleus have been observed in
$^{136}$Nd recently \cite{ Mergel02}.

On the theoretical side, such chiral bands have been predicted by
the particle-rotor model (PRM) and tilted axis cranking (TAC)
model based on the mean field approximation for triaxial deformed
case \cite{Meng97}. The breaking of the chiral symmetry is due to
the fact that the axis of the uniform rotation lies outside any of
the principal plane of the density distribution. The semiclassical
mean field description for tilted nuclear rotation can be traced
back to the 80's in last century \cite{Kerman81, Frisk87,
Frauendorf93}. The qualities of the TAC approximation have been
discussed and tested in Ref. \cite{Meng96} with the PRM model. A
detailed discussion of the self-consistent tilted axis cranking
approach can be found in Ref. \cite{Frauendorf00}. The advantage
of the mean field approach is that it can be easily extended to
the multi-quasiparticle case. However, the mean field approach
violates the rotational invariance and the total angular momentum
is not a good quantum number. The chiral doublet bands are the
results of the TAC in the triaxial case and can be observed
experimentally due to the breaking of the chiral symmetry by
quantum tunnelling effect, as demonstrated in PRM. Although much
effort has been devoted to study the TAC phenomena, the more
microscopic relativistic mean field (RMF) model and Skyrme
Hartree-Fock (SHF) calculations have been reported only in the
context \cite{Madokoro00,Olbratowski02} of planar rotation, i.e.,
magnetic rotation, due to their sophisticated codes and time
consuming numerical procedures. In Ref. \cite{Dimitrov00}, using
hybrid Woods-Saxon and Nilsson model to replace the single
particle energies in the triaxial TAC model of Ref. \cite{Meng97}
and combining with shell correction method, the existence of
chiral characteristic for $^{134}$Pr and $^{188}$Ir has been
demonstrated.

In this paper, the chiral doublet structures for nuclei in
$A\sim100$ and $A\sim130$ regions will be studied with the
particle-rotor model. The rotational levels, $B(M1)$, $B(E2)$, and
the total angular momentum as a function of the rotational
frequency will be presented to analyze the properties of the
chiral doublet bands. The chiral doublet bands observed in
$A\sim130$ region will be described and in $A\sim100$ region will
be predicted.

\section{formulation}

To describe the interplay between the motion of particles and the
collective motion, Bohr and Mottelson proposed to take into
account only a few so-called valence particles, which move more or
less independently in the deformed well of the core, and to couple
them to a collective rotor which stands for the rest of the
particles \cite{AB2}. The division into core and valence particles
is not always unique.  Generally, one divides the Hamiltonian into
two parts: an intrinsic part $H_{intr}$, which describes
microscopically a valence particle or a whole subgroup of
particles near Fermi level; and a phenomenological part $H_{coll}$
which describes the inert core.

The total PRM Hamiltonian is written as: \beq\label{Hamiltonian}
H=H_{intr}+H_{coll}. \eeq Here, we consider the case of a proton
particle (p) and a neutron hole (n) in the intrinsic channel, i.
e., \beq H_{intr}=h_p+h_n. \eeq The formalism can be extended to
the multi-proton or multi-neutron cases.

For a single-j model, the quadrupole deformation potential which
results in the energy level splitting is: \beq
V_p=\frac{206}{A^{1/3}}\beta\ls \cos\gamma
Y_{20}+\frac{\sin\gamma}{\sqrt{2}}\lb Y_{22}+Y_{2-2}\rb\rs\eeq The
corresponding single particle energy for a single-j model can be
obtained as: \beq h_{p(n)}
=\pm\frac{1}{2}C\Lb{\lb{j_{3}^2-\frac{j(j+1)}{3}}\rb
\cos\gamma+\frac{1}{2\sqrt{3}}\ls j^2_++j^2_-\rs \sin\gamma}\Rb,
\eeq where the plus sign refers to a particle, the minus to a hole
and the coupling constant $C$ is: \beq
C=\frac{195}{j(j+1)}A^{-1/3}\beta~{\rm MeV}\eeq

The angular momentum operator of the core $\vec{R}$ and of the
valence particles $\vec{j}$ form the total angular momentum
$\vec{I}$: \beq
   \vec{I}=\vec{R}+\vec{j_p}+\vec{j_n}.
\eeq
For the case of the triaxial rotation,  the moments of
inertia for irrotational flow are
 \beq
  {\cal{J}}_\nu={\cal{J}}\sin^2(\gamma-\frac{2\pi}{3}\nu),~~~~(\nu=1,2,3),
 \eeq
 where ${\cal{J}}$ depends on the quadrupole deformation
$\beta$ and the mass parameter \cite{Meyer75}. Usually the moment
of inertia increases with the angular momentum. For simplicity, a
constant moment of inertia is assumed in the calculation.

The Hamiltonian of the core is \beq
H_{coll}=\sum_{\nu=1}^3\frac{(\hat I_\nu-\hat
j_\nu)^2}{2{\cal{J}}_\nu}~. \eeq
The total Hamiltonian
(\ref{Hamiltonian}) is invariant under $180^\circ$ rotations about
the intrinsic axes ($D_2$ symmetry group). So the eigenvectors of
PRM Hamiltonian can be written as \cite{Davidson}: \beq
|IM\alpha\rangle=\sqrt{\frac{1}{2(1+\delta_{K0})}}\Lb{\sum_{K,k_p,k_n}C_{k_pk_n}^{IK\alpha}\ls
\lr IMKk_pk_n\alpha\rc+(-1)^{I-j_p-j_n}\lr
IM-K-k_p-k_n\alpha\rc\rs}\Rb, \eeq where $|IMK\rangle$ is the
Wigner D-function, $|k_pk_n\rangle$ is the product of the proton
and neutron states $|jk\rangle$, $C_{k_pk_n}^{IK\alpha}$ is the
expansion coefficient and the summation is restricted to $|K|\leq
I$, $|k_i|\leq j_i,~(i=n,p)$, $(K-k_p-k_n)$ even, $k_p+k_n>0$,
when $k_p+k_n=0$, $k_p\geq0$. The angular momentum projections
onto the quantization axis (3-) in the intrinsic frame and the
$z$-axis in the laboratory frame are denoted by K and M,
respectively, and the other quantum numbers are denoted by
$\alpha$.

The reduced transition probabilities are defined as
 \beq
   B(\sigma\lambda, I'\alpha'\to I\alpha)=\sum_{M' M} \lr\lc
    {\rm{f}},~I M\alpha\lr\hat T_{\lambda\mu}\rl
    {\rm{i}},~I'M'\alpha'\rc\rl^2,
 \eeq
 where $\sigma=E$ or $M$ indicates electric or magnetic transition respectively, and
$\lambda$ is the rank of electric or magnetic transition operator
from initial state i to final state f.

The quadrupole moments in the intrinsic system and the laboratory
frame are connected by the relation
 \beq
    \hat Q_{2\mu}={\cal D}^{2*}_{\mu0}\hat Q'_{20}+({\cal D}^{2*}_{\mu2} +
                  {\cal D}^{2*}_{\mu-2})\hat Q'_{22},
 \eeq
 where $\hat Q_{2\mu}$ and $\hat Q'_{2\mu}$ are the transition operators of the core in the
laboratory frame and the intrinsic frame respectively. For the
stretched $E2$ transitions we have
 \beqn
    B(E2,I \alpha \to I' \alpha')
    &=& Q_0^2\frac{5}{16\pi}| \sum^{k_p,k_n}_{K,K'} C_{k_pk_n}^{IK\alpha} C_{k_pk_n}^{I'K'\alpha '}
       \ls \cos\gamma\lc IK20| I' K'\rc\right.\re &&-\frac{\sin\gamma}{\sqrt{2}} \lb\lc
IK22|I' K'\rc\right. +\left.\left.\left. \lc IK2-2|I'
K'\rc\rb\rs\rl^2,
 \eeqn
 in which the contribution from the single particles is neglected as the $Q_0$
value is much large than that from the  single particles.
$Q_0=\frac{3}{\sqrt{5\pi}}R^2_0Z\beta$ is the intrinsic charge
quadrupole momentum, $R_0$ the nuclear radius and $Z$ the charge
number.

The $B(M1)$ values are as the following:
 \beqn &&B(M1,I\alpha \rightarrow
I'\alpha')\re&&=\frac{3}{16\pi}|\sum_{\mu,k_p,k_n,k'_p,k'_n}
\frac{1}{(1+\delta_{K'0})}\frac{1}{(1+\delta_{K0})}C_{k_pk_n}^{IK\alpha}
C_{k'_pk'_n}^{I'K'\alpha '}\re &&\ls\langle{IK1\mu|I'K'}\rangle
\langle{k'_pk'_n|(g_p-g_R)j_{p\mu}+(g_n-g_R)j_{n\mu}|k_pk_n}\rangle\right.\nonumber\\
&&+(-1)^{I-j_p-j_n}\langle{I-K1\mu|I'K'}\rangle
\langle{k'_pk'_n|(g_p-g_R)j_{p\mu}+(g_n-g_R)j_{n\mu}|-k_p-k_n}\rangle
]\re &&+{\rm{sign.}}|^2\eeqn where sign. represents the
contribution due to the symmetry of state $\lr IM\alpha\rc$ in Eq.
(9), which is the same as the first two terms in Eq. (13) by the
replacement $K'\to-K', ~~k'_p\to-k'_p$ and $k'_n\to-k'_n$. The
spherical tensor of rank 1 $j_\mu$ is: \beq
j_\mu=\lb{j_0=j_3,j_{\pm 1}=\frac{\mp(j_1\pm ij_2)}{\sqrt{2}}}\rb.
\eeq

\section{ results and Discussions}
For the numerical calculation, we follow the procedures in Ref.
\cite{Meng97} to study the chiral doublet structures in $A\sim100$
and $A\sim130$ nuclei using the particle-rotor model. Firstly, the
rotational spectra of these four isotones in $A\sim130$ nuclei in
Ref. \cite{Starosta01} are calculated and compared with the
experimental data.  We fix the configuration as $\pi
h_{11/2}\otimes\nu h_{11/2}^{-1}$ and ${\cal J}=25$ MeV$^{-1}$.
The parameters in our calculation are listed in Table I. Secondly,
the proton and neutron configuration $\pi g_{9/2}\otimes\nu
g_{9/2}^{-1}$ is adopted in the calculation of $A\sim100$ nuclei
with ${\cal J}=30$ MeV$^{-1}$, $\gamma=-30^\circ$, and $C=0.1$
MeV, $0.20$ MeV, $0.25$ MeV corresponding to the deformation
$\beta\approx0.06$, $0.12$, $0.15$, respectively. In the
calculation of $B(M1)$ and $B(E2)$, the quadrupole moments
$Q'_{21}$ and $Q'_{2-1}$ are zero in the intrinsic fame by
definition. Since we are only interested in the trends of the
$B(M1)$ and $B(E2)$ instead of the absolute values of $B(M1)$ and
$B(E2)$, the g-factors $g_{p(n)}$$-$$g_R$ are set as $1$ or $-1$
for the proton or neutron, respectively, and the intrinsic
quadrupole moments are chosen as $Q'_{20}=\cos\gamma$,
$Q'_{22}=Q'_{2-2}=-\sin\gamma/\sqrt{2}$. Finally, the chiral
doublet structures with asymmetric particle-hole configuration
$\pi g_{9/2}^{-1}\otimes \nu h_{11/2}$ in $^{104}$Rh are also
discussed.

\begin{figure}[htbp]
 \centering
 \includegraphics[height=12cm,width=10cm,angle=-90]{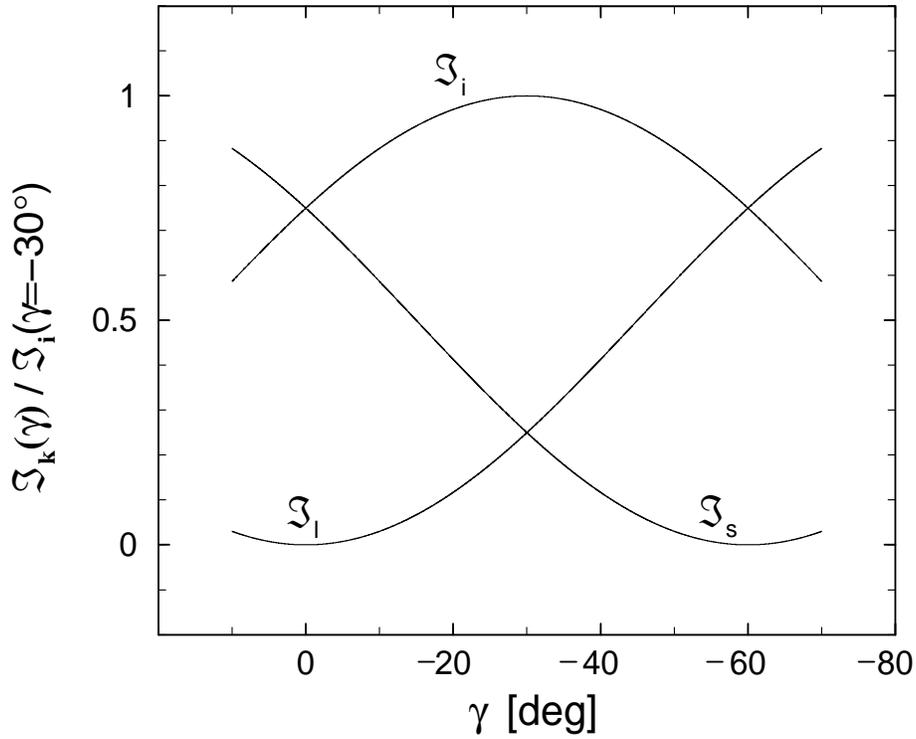}
 \caption{The moments of inertia for irrotational flow as functions of $\gamma$ deformation.}
 \label{fig:ROI}
\end{figure}

For $\gamma=-30^\circ$, we have  the axial length relation
$R_1<R_2<R_3$ and denote the intrinsic axes 1, 2, 3 as the s, i, l
axes, respectively \cite{AB2}. The moment of inertia as a function
of $\gamma-$deformation is presented in Fig. \ref{fig:ROI}. When
$\gamma=-30^\circ$, one has ${\cal J}_l={\cal
J}_s=\dfrac{1}{4}{\cal J}_i$, i.e., the collective angular
momentum vector $\vec R$ tends to align along the intermediate
axis, which minimizes the rotational energy \cite{AB2}. The single
particle and hole angular momentum vectors will tend to align
along the short and long nuclear axis, respectively. These three
angular momenta are mutually perpendicular to each other. These
orientations maximize the overlap of the particle densities with
the triaxial potential, and result in minimizing the interaction
energy \cite{Meng97}.

\begin{figure}[htbp]
 \centering
 \includegraphics[height=12cm,width=10cm,angle=-90]{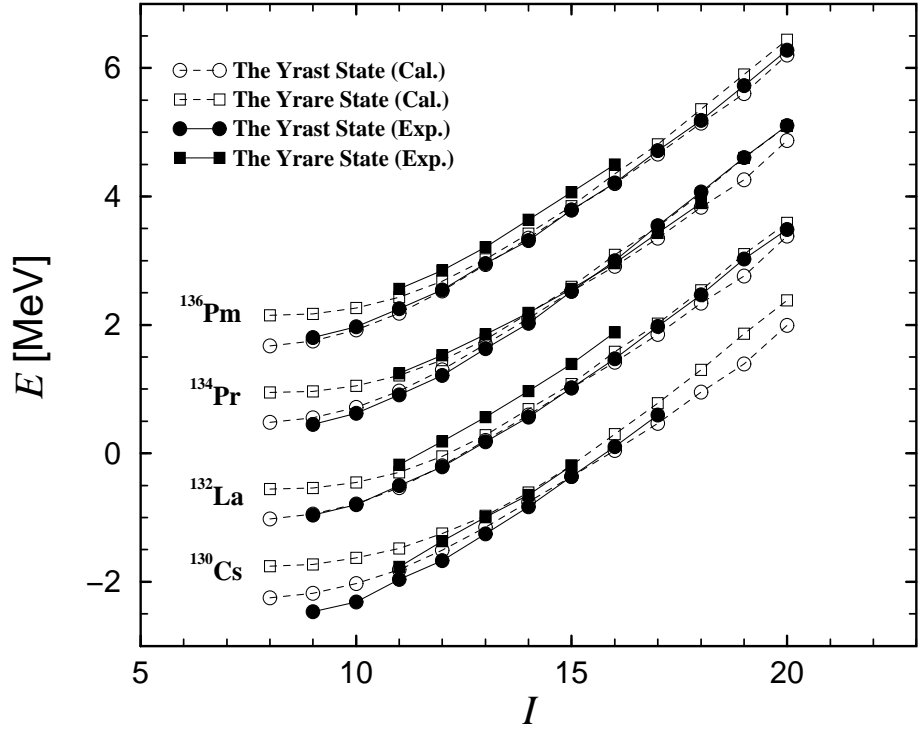}
 \caption{Calculated and experimental energies for yrast
  band (circles) and yrare band (squares) with the configuration
  of $\pi h_{11/2}\otimes\nu h_{11/2}^{-1}$ in four $N=75$
  isotones $^{130}$Cs, $^{132}$La, $^{134}$Pr, $^{136}$Pm.
  The open symbols correspond to the calculated values.
  The filled symbols correspond to the experimental values.
 For the details, see the text.}
 \label{fig:PJbeta}
\end{figure}

The four pairs of chiral doublet bands have been observed in
$A\sim130$ region \cite{Starosta01}. The observed and the
calculated energy spectra for $^{130}$Cs, $^{132}$La, $^{134}$Pr,
$^{136}$Pm are presented in Fig. \ref{fig:PJbeta}. The
configuration $\pi h_{11/2}\otimes\nu h_{11/2}^{-1}$ and the
moment of inertia of the rotor ${\cal J}=$ 25 MeV$^{-1}$ have been
chosen in the calculation. The available data and also the
recommended bandhead spin ($I=9\hbar$) are taken from Ref.
\cite{Starosta01}. According to the $\varepsilon$ deformation of
3D TAC calculation \cite{Starosta01}, the corresponding parameters
$C$ are chosen as in Table I. Similar as in Ref.
\cite{Starosta01}, the calculated spectra are separated by 1.5 MeV
for display. The data of $^{130}$Cs, $^{132}$La, $^{134}$Pr,
$^{136}$Pm are shifted by $-2.47$ MeV, $-0.96$ MeV, $+0.45$ MeV
and $+1.80$ MeV respectively. In such a way, the calculated
energies are set to be equal to the experimental data at
$I=15\hbar$, so that the experimental and the calculated spectra
can be compared easily. In Fig. \ref{fig:PJbeta}, the calculated
and observed energy levels are coincident with each other and both
show the energy degeneracy around $I=15\hbar$ for $^{134}$Pr. For
the other $N=75$ isotones, the sidebands are displaced modestly in
energy from the yrast bands. For $^{130}$Cs, this displacement
stays roughly constant at $\approx0.22$ MeV in the data and at
$\approx0.25$ MeV in the calculated data, respectively, in the
spin interval $11\hbar<I<15\hbar$. There is a relatively large
energy displacement between chiral bands in $^{132}$La and
$^{136}$Pm. It is shown that the chiral doublet bands observed in
$A\sim130$ nuclei can be well understood from the PRM similarly as
in the Refs. \cite{Starosta02,StarostaNPA01}. The mutual
orientation of angular momentum components has also been studied
and the same conclusion as Refs. \cite{Starosta02,StarostaNPA01}
has been drawn, namely, the spontaneous chiral symmetry breaking
leads to the doubling of states.

After the discussion for $A\sim130$ nuclei, we wonder whether the
chiral doublets exist in $A\sim100$ nuclei with the configuration
$\pi g_{9/2}\otimes\nu g_{9/2}^{-1}$.

\begin{figure}[htbp]
 \centering
 \includegraphics[height=15cm,width=8cm,angle=-90]{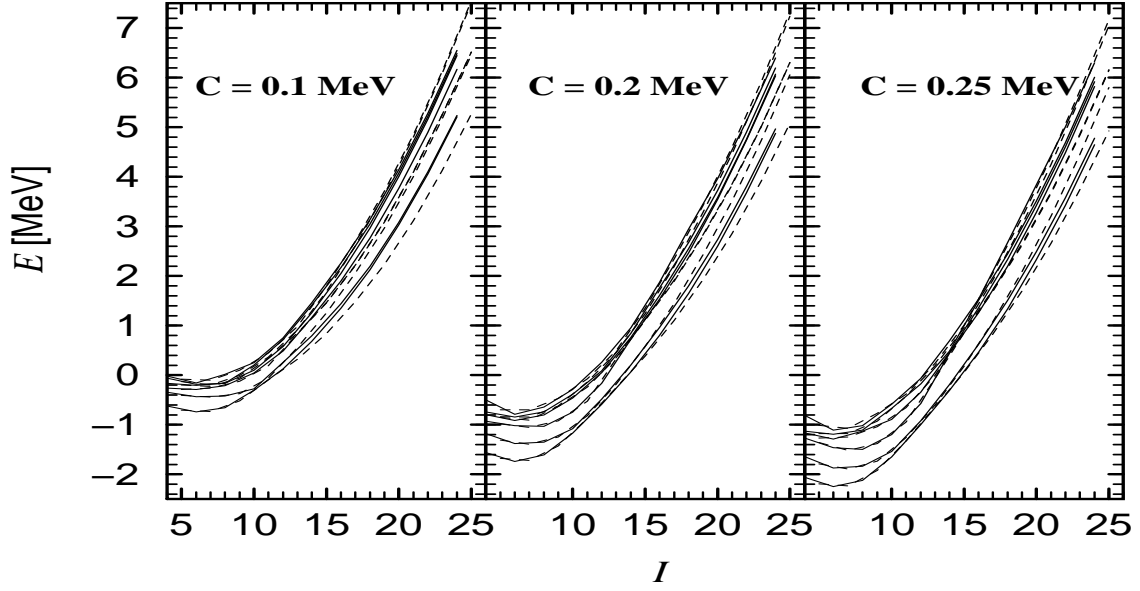}
 \caption{Rotational energies versus total angular momentum for the configuration $\pi g_{9/2}\otimes\nu g_{9/2}^{-1}$
  calculated with ${\cal J}=30$ MeV$^{-1}$, $\gamma=-30^\circ$ and different $C$ values.
 Full lines correspond to even and dashed to odd spin. }
 \label{fig:EI}
\end{figure}

For the $A\sim100$ nuclei, the rotational bands obtained from the
PRM Hamiltonian with ${\cal J} = 30$ MeV$^{-1}$ and $\gamma =
-30^\circ$, are presented in Fig. \ref{fig:EI}. The left, middle,
and right panel corresponds to the case of $C=0.1$ MeV, $0.2$ MeV,
and $0.25$ MeV, respectively. The combination of the proton
particle with the neutron hole favors the s-l plane. At the
beginning of the bands, the collective angular momentum $\vec R$
is small and the total angular momentum $\vec I$ lies in a
principal plane defined by the short and long principal axes of
the deformed nucleus. There are two degenerate solutions (planar
solution) obtained by the rotation $R_3(\pi)$ (or $R_1(\pi)$) that
can be combined into degenerate states of opposite signature,
similar as in Ref. \cite{Meng97}. In all the panels of Fig.
\ref{fig:EI}, the characteristic of $\triangle I=1$ bands appears
in low lying states before $I=11\hbar$. With the spin increasing,
$\vec R$ becomes comparable to the single particle angular
momentum, $\vec I$ gradually turns towards the i-axis and does not
lie in any of the principal plane. This is demonstrated in all of
the three panels of Fig. \ref{fig:EI} with the two lowest bands
becoming nearly degenerate around $I=12\hbar$. In the left panel,
the yrast band and the yrare band are close to each other ear
$I=12\hbar$ despite the fact that the overall energy displacement
between the two lowest bands is relatively large. In the middle
panel, the two lowest bands become nearly degenerate near the spin
interval $11\hbar<I<13\hbar$. The interval of energy degeneracy is
$11\hbar<I<15\hbar$ in the right panel. This shows that the chiral
symmetry favors the case with larger deformation. After
$I=15\hbar$, the degenerate energy band splits into four bands
with $\bigtriangleup I=2$.  The near degeneracy for the pairs of
$\bigtriangleup I=1$ bands give the signal that chiral doublet
bands exist in $A\sim100$ nuclei with the configuration $\pi
g_{9/2}\otimes\nu g_{9/2}^{-1}$.

\begin{figure}[htbp]
 \centering
 \includegraphics[height=14cm,width=8cm,angle=-90]{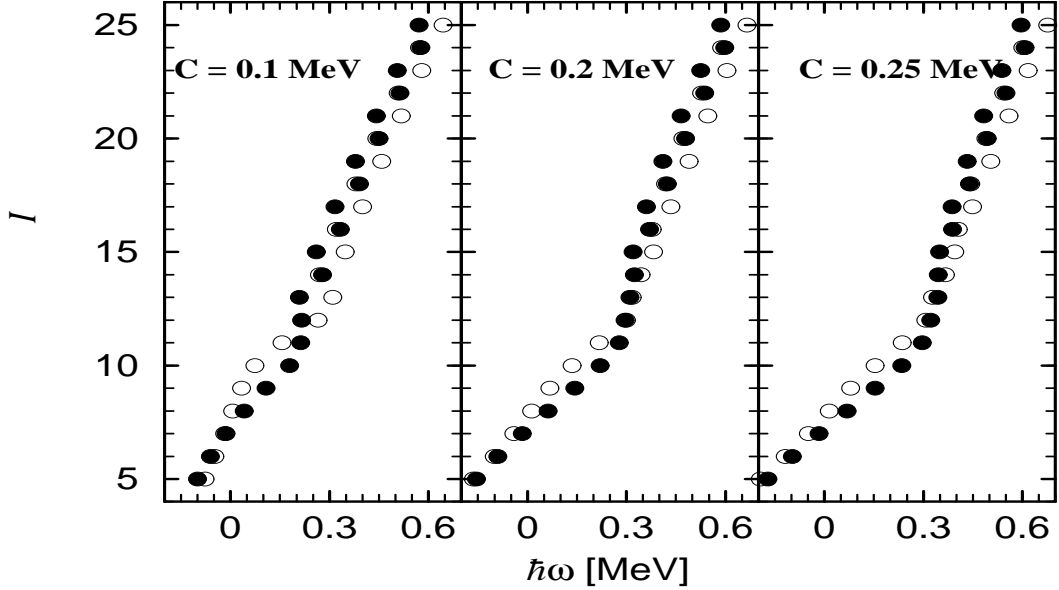}
 \caption{The total angular momentum as a function of the rotational frequency for the configuration $\pi g_{9/2}\otimes\nu g_{9/2}^{-1}$
 calculated with $\gamma=-30^\circ$ and ${\cal{J}}=30$ Mev$^{-1}$.
  The filled and open circles represent the yrast and the yrare band, respectively.}
 \label{fig:IW}
\end{figure}

From the PRM energies in Fig. \ref{fig:EI}, the frequency can be
calculated by means of $\omega(I)=(E(I+1)-E(I-1))/2$
\cite{Meng96}. The functions $I(\omega)$ are presented in Fig.
\ref{fig:IW}. The deformation parameter $C$ in the left, middle
and right panel is taken as $0.1$ MeV, $0.2$ MeV, and $0.25$ MeV,
respectively, as in Fig. \ref{fig:EI}. In Fig. \ref{fig:IW}, when
$I<8\hbar$, the frequency $\omega$ is negative because the total
angular momentum near the bandhead is mainly provided by the
orientation of the particle and hole. This is the region where the
classical mean field approximation does not work. The negative
rotational frequency means that rotation is not the collective
rotation usually assumed in cranking approximation but intrinsic
excitations. When $8\hbar \le I \le 12 \hbar$ ( $0 \le \hbar\omega
\le 0.3$ MeV ), the total angular momentum is provided by the
single particle and hole together with the core. The rotation is
aplanar. The bands in this spin region show up as a nearly
straight line in all three panels of Fig. \ref{fig:IW}. For higher
spins, the core becomes dominant to contribute to the total
angular momentum and the angular momentum will align along the
i-axis leading to the kink of the second kind of the moment of
inertia ${\cal {J}}^{(2)}=dI (\omega)/d \omega$ (see in Fig.
\ref{fig:IW}), which is the slope of the curve $I(\omega)$. This
kink represents a reorientation of the total angular momentum from
the aplanar towards the i-axis. After the kink, the angular
momenta of the core and the single particles will align along the
intermediate axis which is the direction of the total angular
momentum and the signature splitting appears. It should be noted
that the stability of chiral geometry depends on the competition
between strength of the single particle Hamiltonian and the
Coriolis force whose strength is related to the moment of inertia.
The signature splitting should not appear if chiral geometry is
always stable.

\begin{figure}[htbp]
 \centering
 \includegraphics[height=12cm,width=16cm]{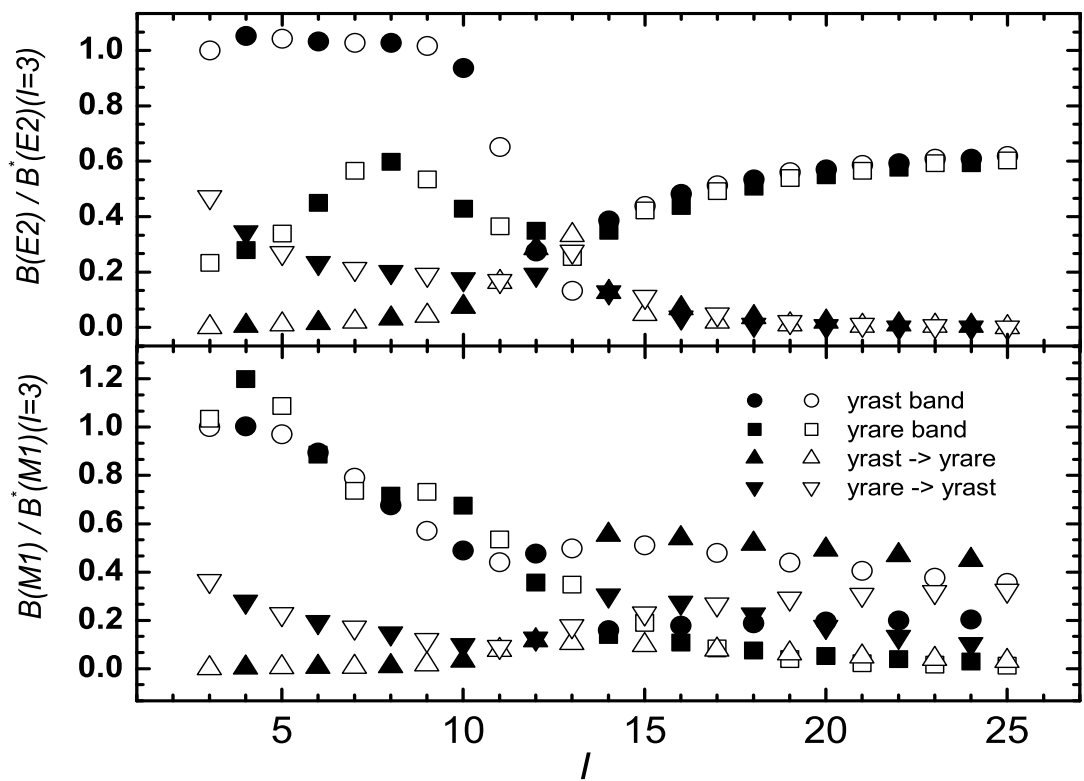}
 \caption{$B(M1)$ and $B(E2)$ values as functions of the total angular momentum for a $g_{9/2}$
 particle and a $g_{9/2}$ hole coupled to a triaxial rotor with $\gamma=-30^\circ$.
$B^*(M1)(I=3)$ and $B^*(E2)(I=3)$ are respectively the intra
$B(M1)$ and $B(E2)$ transitions of the yrast band at $I=3\hbar$.
The filled symbols correspond to the even spin, while the open
ones to the odd spin. The symbols, circles, squares, triangle ups
and triangle downs represent respectively the intra transitions of
the yrast band, the intra transitions of the yrare band, the inter
transitions from the yrast band to the yrare band and the inter
transitions from the yrare band to the yrast band.  }
 \label{fig:newBME}
\end{figure}

In order to study chiral doublet structures, the reduced $B(M1)$
and $B(E2)$ transition probabilities are presented in Fig.
\ref{fig:newBME} for the case of $C=0.2$ MeV and ${\cal{J}}=30$
MeV$^{-1}$. Since we are only interested in the trends of the
$B(M1)$ and $B(E2)$, the transition values in Fig.
\ref{fig:newBME} are given in a units of the $B(M1)$ and $B(E2)$
yrast intraband transitions at $I=3\hbar$ respectively. The upper
panel shows the $B(E2)$ transition probabilities, and the lower
panel corresponds to the $B(M1)$ values. The filled symbols and
the open symbols represent the transition probabilities of the
bands with even spin and odd spin, respectively. The inter band
transition probabilities reflect the change from the planar to the
aplanar rotation \cite{Meng97}. At low spins, it is the case of
planar rotation, and there are weak inter band transition but
strong intra band transitions in Fig. \ref{fig:newBME}. Near
$I=12\hbar$, where the two bands come very close and the rotation
becomes aplanar, the inter band transitions in both directions are
observed. At the crossing point, the two intrinsic structures are
mixed, and the inter $B(E2)$ transition is larger than the intra
$B(E2)$ transition. After $I=15\hbar$, the rotation rapidly aligns
towards the principal i-axis. There appears the spin-dependent
splitting in the level energies. The $B(E2)$ transition isn't
affected by this splitting. This is a regular electric transition
between $\bigtriangleup I=2$ bands as shown in the upper panel of
the Fig. \ref{fig:newBME}. Moreover, it is observed that the inter
band $B(E2)$ transitions disappear. After $I=15\hbar$, the
calculated $B(M1)$ transition in the lower panel of the Fig.
\ref{fig:newBME} show odd-even staggering. The dominant $B(M1)$
transition appears for the intra band of the yrast band with odd
spin and the inter band (yrast band to yrare band) with even spin.
The above conclusions are consistent with $B(M1)/B(E2)$ staggering
discussed for the partner bands in Refs. \cite{Koike03, KoikeAIP}.

\begin{figure}[htbp]
 \centering
 \includegraphics[height=13cm,width=9.5cm,angle=-90]{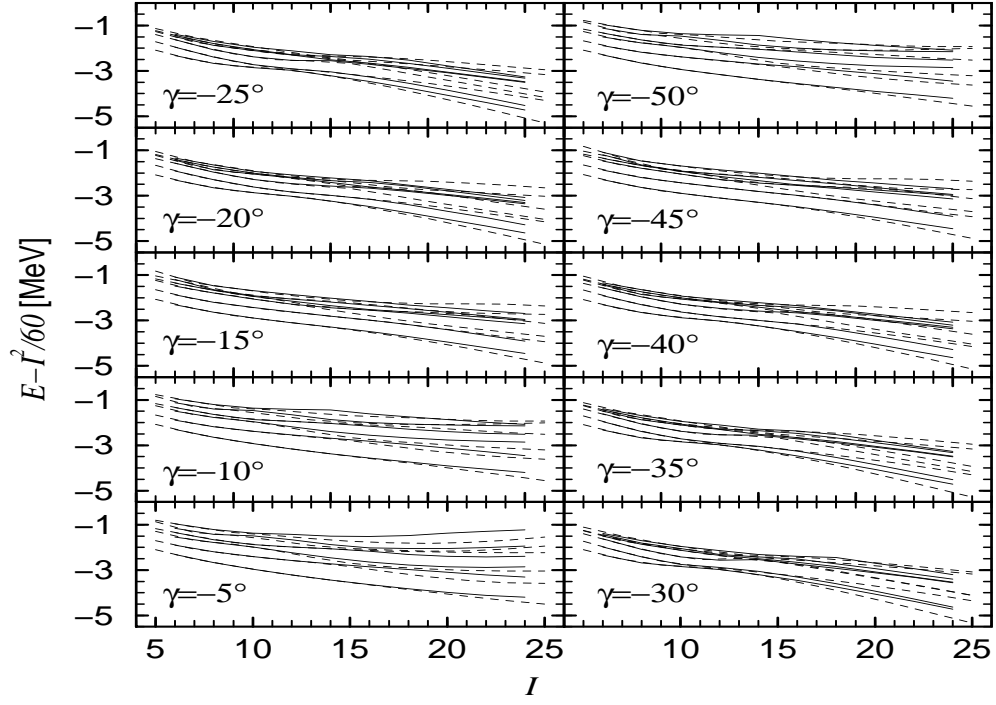}
 \caption{Rotational bands of a $g_{9/2}$ proton particle and a $g_{9/2}$ neutron hole coupled
 to a triaxial rotor with $C=0.2$ MeV and ${\cal{J}}=30$ MeV$^{-1}$. Different values of $\gamma$ ($-5^\circ$,
 $-10^\circ$, $\cdots$, $-50^\circ$) are taken in the calculation. Full lines correspond to even and dashed to odd spin. }
 \label{fig:EnewI2}
\end{figure}

To find out the range of $\gamma-$deformation for the appearance
of chiral doublet bands, the calculated level energies for a
particle and a hole coupled to the triaxial rotor with different
$\gamma$ values are shown in Fig. \ref{fig:EnewI2}, similar as in
Ref.\cite{Meng97, Starosta02} but for different configuration. We
find that the chiral doublet bands appear in the interval
$-35^\circ < \gamma < -25^\circ$. It should be noted that the
difference between yrast bands and yrare bands has been amplified
by subtracting a rotor energy $I^2/60$ MeV. In most cases, the
energy difference is within 0.7 MeV. It indicates there is a
certain margin of the $\gamma-$deformation, where the energy
degeneracy is expected.  The best condition for the appearance of
the chiral doublet bands is $\gamma=-30^\circ$ in $A\sim100$
region with the configuration of $\pi g_{9/2}\otimes\nu
g_{9/2}^{-1}$, similar as in Ref.\cite{Meng97, Starosta02}.

\begin{figure}[htbp]
 \centering
 \includegraphics[height=13cm,width=9.5cm,angle=-90]{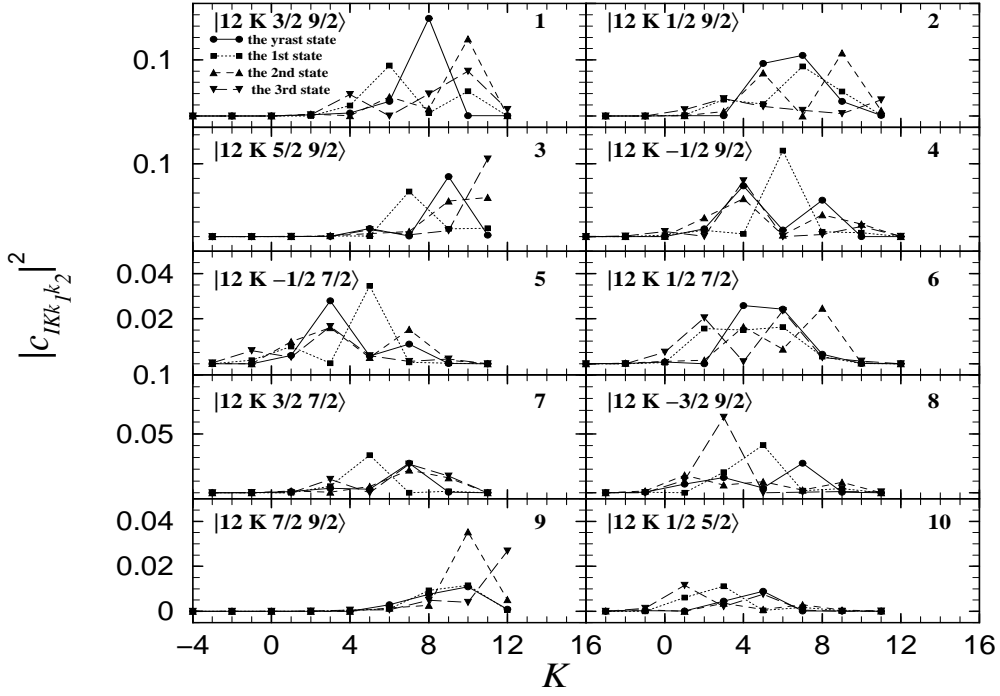}
 \caption{The main structure of the triaxial PRM eigenstates as
 a function of the total projection K for $C=0.25$ MeV, $\gamma=-30^\circ$ and $I=12\hbar$.
 The probabilities of the yrast band, the yrare band, the second excited band
 and the third excited band are given  by filled circle, filled square, filled
 triangle up, and down symbols, respectively. The quantum numbers for corresponding basis states are indicated
 in the upper left corner of each panel.}
 \label{fig:ckI1225}
\end{figure}

To further investigate the microscopic reasons of chiral doublet
bands, the structure of the wave functions for the yrast bands and
the yrare bands for the configuration $\pi g_{9/2}\otimes\nu
g_{9/2}^{-1}$ are presented in Fig. \ref{fig:ckI1225}. They are
taken from the triaxial PRM calculation with $C=0.25$ MeV,
${\cal{J}}=30$ MeV$^{-1}$ and $\gamma=-30^\circ$. Here, only the
first term in Eq.(9) is adopted to calculate the probability
distribution as the function of K. We choose the top ten basis
states $\lr 12MKk_1k_2\rc$ with the largest  contribution to the
yrast band states. The probability distributions are far more
complicated than we have expected. This complexity indicates that
the composition of the total angular momentum is not as simple as
in the picture of the classical mean field approximation and the
quantum effects are very important; Moreover, this complexity may
be significantly reduced with different choice of basis states
\cite{Starosta02}. In Fig. \ref{fig:ckI1225}, the yrare bands
(filled squares) have the similar probability distributions with
the yrast bands (filled circles) in many panels. It partly
provides the reason why there is energy degeneracy between these
two bands. Meanwhile, the quantum effect is so important that the
chiral symmetry is broken, as can be seen also in the $B(M1)$ and
$B(E2)$ values in Fig. \ref{fig:newBME}.

\begin{figure}[htbp]
 \centering
 \includegraphics[totalheight=12cm,angle=-90]{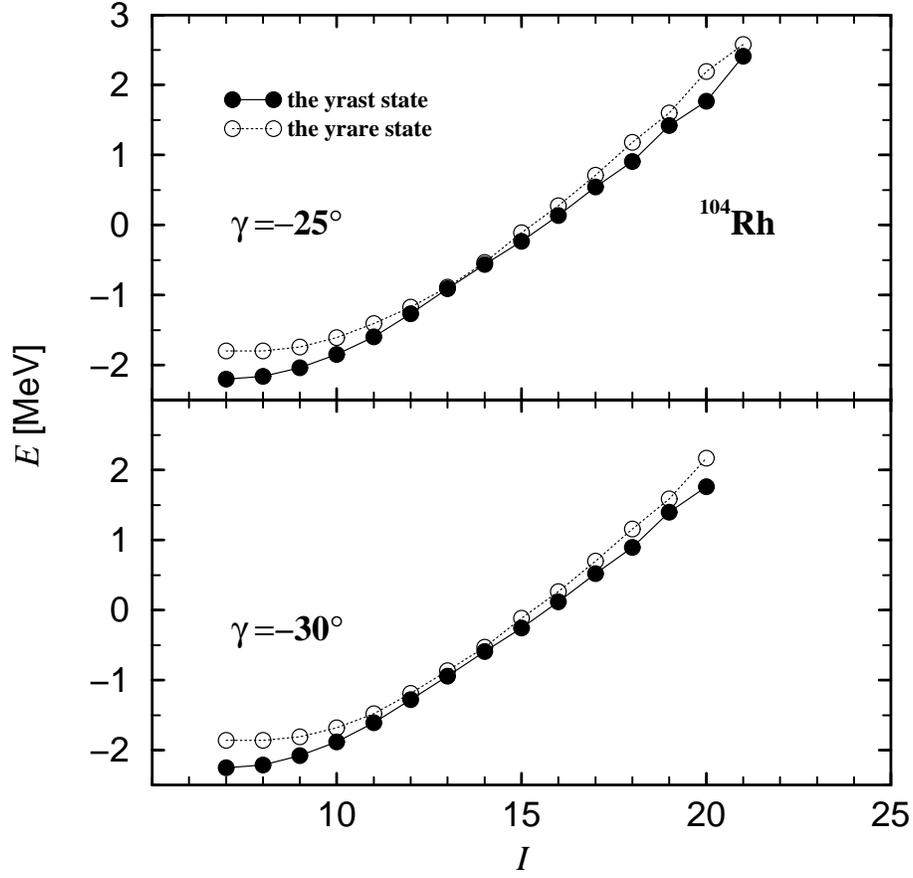}
 \caption{Rotational spectra for the yrast and yrare bands with the configuration
 $\pi g_{9/2}^{-1}\otimes \nu h_{11/2}$, $C=0.2$ MeV and ${\cal{J}}=30$ MeV$^{-1}$.
 The upper and lower panel shows the case of $\gamma=-25^\circ$ and $\gamma=-30^\circ$, respectively. The filled circles indicate the yrast band, and the
 open circles indicate the yrare band.}
 \label{fig:Rh1043}
\end{figure}

Recently, Koike $et~al$. \cite{KoikeAIP} reported the chiral
doublets in $^{104}$Rh with the asymmetrical configuration $\pi
g_{9/2}^{-1}\otimes\nu h_{11/2}$, which motivated the study of the
chiral symmetry breaking in $^{104}$Rh here. The energy spectra
for yrast and yrare bands with the configuration $\pi
g_{9/2}^{-1}\otimes \nu h_{11/2}$, ${\cal{J}}=30$ MeV$^{-1}$ and
$C=0.2$ MeV  are presented in Fig. \ref{fig:Rh1043}. The
corresponding nuclei with such configuration can be found easily
in $A\sim100$ region, e.g., $^{104}$Rh. The upper and lower panel
of Fig. \ref{fig:Rh1043} shows the calculatd energies as a
function of spin for $\gamma=-25^\circ$ and $\gamma=-30^\circ$,
respectively. In both panels, the two lowest bands are near
degenerate in the spin interval $11\hbar<I<16\hbar$. It shows that
the chiral doublets exist also in nuclei with asymmetric
particle-hole configurations.

\section{conclusions}
The chiral doublets for nuclei in $A\sim100$ and $A\sim130$
regions have been studied with the particle-rotor model. The
experimental spectra of chiral partners bands for $N=75$ isotones
$^{130}$Cs, $^{132}$La, $^{134}$Pr, $^{136}$Pm in $A\sim130$
region  \cite{Starosta01} have been well reproduced by the
calculation with the configuration $\pi h_{11/2}\otimes\nu
h_{11/2}^{-1}$. The possible chiral doublets in $A\sim100$ region
corresponding to a $\gamma-$deformation in the interval $-35^\circ
< \gamma < -25^\circ$ have been predicted by the PRM model with
the configuration $\pi g_{9/2}\otimes\nu g_{9/2}^{-1}$ based on
the analysis of the spectra, the $\omega$-I relation, the $B(M1)$
and $B(E2)$ transition probabilities, and the structure of the
wave functions. The possible chiral doublets for nuclei with
asymmetric particle and hole configuration are also discussed.

\begin{acknowledgments}

This work is partly supported by the Major State Basic Research
Development Program Under Contract Number G2000077407 and the
National Natural Science Foundation of China under Grant No.
10025522, 10047001 and 19935030.

\end{acknowledgments}

\begin{table}[h]
\centering \caption{The parameters in the triaxial PRM
calculation. Compared with Ref. \cite{Starosta01}, the same
$\gamma$ value is taken and the parameter $C$ is comparable with
$\varepsilon$.($\varepsilon\approx0.95\beta$) }
\label{tab:paramet} \btab{c c c c c c }\hline\hline
        ~~~~~~&~~~$^{130}_{75}$Cs~~~&~~~ $^{132}_{75}$La~~~&~~~ $^{134}_{75}$Pr~~~&~~~ $^{136}_{75}$Pm~~~ \\ \hline
      $C$ [MeV]~~&~~~0.175 ~~~&~~~0.19 ~~~&~~~0.19 ~~~&~~~0.21 ~~~\\ \hline
      $\varepsilon$~\cite{Starosta01}~~&~~~0.16 ~~~&~~~0.175 ~~~&~~~0.175 ~~~&~~~0.195 ~~~\\ \hline
      $\gamma$  ~\cite{Starosta01}~~    &~~~$-39$  ~~~ &~~~$-32$   ~~~ &~~~$-27$   ~~ ~&~~~$-27$   ~ ~~\\ \hline
\hline \etab
\end{table}

\end{document}